\documentclass{article}

     \usepackage[final]{neurips_2020}

\usepackage[utf8]{inputenc} 
\usepackage[T1]{fontenc}    
\usepackage{hyperref}       
\usepackage{url}            
\usepackage{booktabs}       
\usepackage{amsfonts}       
\usepackage{nicefrac}       
\usepackage{microtype}      
\usepackage{multirow}
\usepackage{color,soul}

\usepackage{tabularx}
\usepackage{threeparttable}
\usepackage{booktabs}
\usepackage{bm}
\usepackage{amsmath}
\usepackage{stfloats}
\usepackage{gensymb}
\usepackage{graphicx}
\usepackage[square,numbers,sort&compress]{natbib}

\title{Encoding Clinical Priori in 3D Convolutional Neural Networks for Prostate Cancer Detection in bpMRI}

\author{
  Anindo Saha, Matin Hosseinzadeh, Henkjan Huisman                              \\
  Diagnostic Image Analysis Group, Radboud University Medical Center            \\
  Nijmegen 6525 GA, The Netherlands                                             \\
  \texttt{\{anindya.shaha,matin.hosseinzadeh,henkjan.huisman\}@radboudumc.nl}   \\
}

\begin{document}

\maketitle

\begin{abstract}
    We hypothesize that anatomical priors can be viable mediums to infuse domain-specific clinical knowledge into state-of-the-art convolutional neural networks (CNN) based on the U-Net architecture. We introduce a probabilistic population prior which captures the spatial prevalence and zonal distinction of clinically significant prostate cancer (csPCa), in order to improve its computer-aided detection (CAD) in bi-parametric MR imaging (bpMRI). To evaluate performance, we train 3D adaptations of the U-Net, U-SEResNet, UNet++ and Attention U-Net using 800 institutional training-validation scans, paired with radiologically-estimated annotations and our computed prior. For 200 independent testing bpMRI scans with histologically-confirmed delineations of csPCa, our proposed method of encoding clinical priori demonstrates a strong ability to improve patient-based diagnosis (upto 8.70\% increase in AUROC) and lesion-level detection (average increase of 1.08 pAUC between 0.1--10 false positives per patient) across all four architectures.
\end{abstract}

\section{Introduction}
\label{intro}

State-of-the-art CNN architectures are often conceived as \textit{one-size-fits-all} solutions to computer vision challenges, where objects can belong to one of 1000 different classes and occupy any part of natural color images \cite{ImageNet}. In contrast, medical imaging modalities in radiology and nuclear medicine exhibit much lower inter-sample variability, where the spatial content of a scan is limited by the underlying imaging protocols and human anatomy. In agreement with recent studies \cite{AnatomicalPriors,WeakProstate,ZonalPrior}, we hypothesize that variant architectures of U-Net can exploit this property via an explicit anatomical prior, particularly at the task of csPCa detection in bpMRI. To this end, we present a probabilistic population prior $P$, constructed using radiologically-estimated csPCa annotations and CNN-generated prostate zonal segmentations of 700 training samples. We propose $P$ as a powerful means of encoding clinical priori to improve patient-based diagnosis and lesion-level detection on histologically-confirmed cases. We evaluate its efficacy across a range of popular 3D U-Net architectures that are widely adapted for biomedical applications \cite{3DUNet,SE,UNet++,Atn-Gates,USENet}.

\paragraph{Related Work} Traditional image analysis techniques, such as MALF \cite{MALF}, can benefit from spatial priori in the form of \textit{atlases} or multi-expert labeled template images reflecting the target organ anatomy. Meanwhile, machine learning models can adapt several techniques, such as reference coordinate systems \cite{PolarBreast,SpectralBrain} or anatomical maps \cite{AnatomicalPriors}, to integrate domain-specific priori into CNN architectures. In recent years, the inclusion of zonal priors \cite{ZonalPrior} and prevalence maps \cite{WeakProstate} have yielded similar benefits in 2D CAD systems for prostate cancer.

\begin{figure}[t!]
\centering
\includegraphics[width=1.000\textwidth]{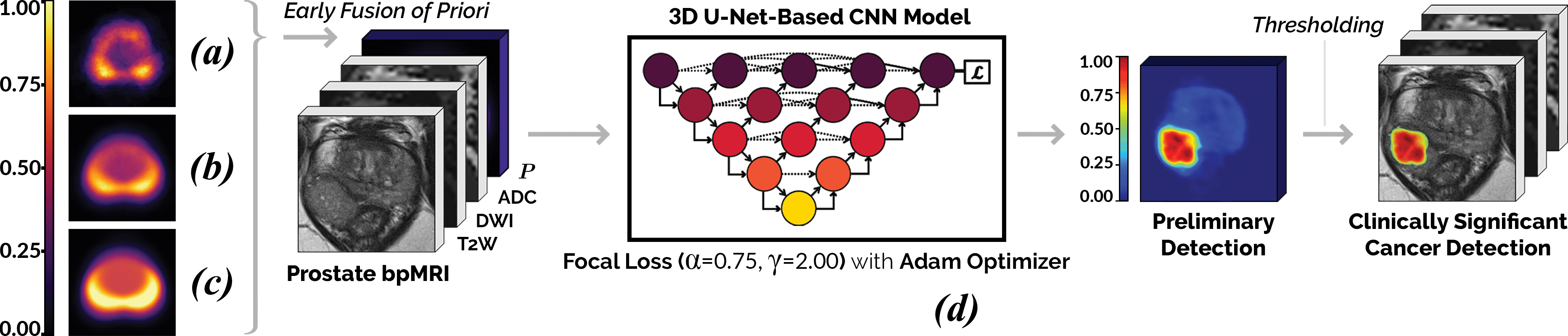}
\caption{\textbf{\textit{(a)}} \textbf{Prevalence Prior}: \textit{P} at $\mu=0.00$ is equivalent to the mean csPCa annotation in the training dataset; mapping the common sizes, shapes and locations of \textit{malignant} lesions. \textbf{\textit{(b)}} \textbf{Hybrid Prior}: \textit{P} at $\mu=0.01$ blends the information of csPCa annotations with that of the prostate zonal segmentations. \textbf{\textit{(c)}} \textbf{Zonal Prior}: \textit{P} at $\mu=0.33$ is approximately equivalent to the weighted average of all prostate zonal segmentations in the training dataset. \textbf{\textit{(d)}}: Schematic of the pipeline used to train/evaluate each candidate 3D CNN model with a variant of the prior $P$, in separate turns.}
\label{fig1}
\end{figure}

\paragraph{Anatomical Priors} For the $i$-th bpMRI scan in the training dataset, let us define its specific prevalence map as $p_i = ({p_i}^1, {p_i}^2,..., {p_i}^n)$, where $n$ represents the total number of voxels per channel. Let us define the binary masks for the prostatic transitional zone (TZ), peripheral zone (PZ) and malignancy (M), if present, in this sample as $B_{TZ}$, $B_{PZ}$ and $B_M$, respectively. We can compute the value of the $j$-th voxel in $p_i$ as follows:

{\setlength{\abovedisplayskip}{-3pt}
\setlength{\belowdisplayskip}{1pt}
\begin{equation*}
\begin{split}
f({p_i}^j) = \begin{cases}
      0.00         & {p_i}^j \in                (B_{TZ} \cup B_{TZ} \cup {B_{M}})' \\
       \mu         & {p_i}^j \in \hspace{3.5pt}  B_{TZ} \cap {B_{M}}' \\
      3\mu         & {p_i}^j \in \hspace{3.5pt}  B_{PZ} \cap {B_{M}}'  \\
      1.00         & {p_i}^j \in \hspace{3.0pt}  B_{M} 
\end{cases}
\end{split}
\end{equation*}}

\noindent Here, $f({p_i}^j)$ aims to model the spatial likelihood of csPCa by drawing upon the empirical distribution of the training dataset. Nearly 75\% and 25\% of all \textit{malignant} lesions emerge from PZ and TZ, respectively \citep{mpMRI,csPCaZones}. Thus, similar to PI-RADS v2 \cite{PIRADSv2}, $f({p_i}^j)$ incorporates the importance of zonal distinction during the assessment of csPCa. In terms of the likelihood of carrying csPCa, it assumes that voxels belonging to the background class are not likely ($f({p_i}^j)=0.00$), those belonging to TZ are more likely ($f({p_i}^j)=\mu$), those belonging to PZ are three times as likely as TZ ($f({p_i}^j)=3\mu$), and those containing csPCa are the most likely ($f({p_i}^j)=1.00$), in any given scan. All the computed specific prevalence maps can be generalized to a single probabilistic population prior, $P = (\sum {p_i})/N \in [0,1]$, where $N$ represents the total number of training samples. The value of $\mu \in [0,0.33]$ is a hyperparameter that regulates the relative contribution of \textit{benign} prostatic regions in the composition of each $p_i$ and subsequently our proposed prior $P$ (refer to  \hyperref[fig1]{Fig. 1(\textit{a-c}})). Due to the standardized bpMRI imaging protocol \cite{PIRADSv2}, inter-sample alignment of the prostate gland is effectively preserved with minimal spatial shifts observed across different patient scans. Prior-to-image correspondence is established at both train-time and inference by using the case-specific prostate segmentations to translate, orient and scale $P$, accordingly, for each bpMRI scan. No additional non-rigid registration techniques have been applied throughout this process.

\section{Experimental Analysis}
\label{exp}
\paragraph{Materials} To train and tune each model, we use 800 prostate bpMRI (T2W, high \textit{b}-value DWI, computed ADC) scans from Radboud University Medical Center, paired with fully delineated annotations of csPCa. Annotations are estimated by a consensus of expert radiologists via PI-RADS v2 \cite{PIRADSv2}, where any lesion marked PI-RADS $\geq$ 4 constitutes as csPCa. From here, 700 and 100 patient scans are partitioned into training and validation sets, respectively, via stratified sampling. To evaluate performance, we use 200 testing scans from Ziekenhuisgroep Twente. Here, annotations are clinically confirmed by independent pathologists \cite{ZGT,ISUP} with Gleason Score $>3+3$ corresponding to csPCa. TZ, PZ segmentations are generated for every scan using a multi-planar, anisotropic 3D U-Net from a separate study \citep{3DSegMod}, where the network achieves an average Dice Similarity Coefficient of $0.90\pm0.01$ for whole-gland segmentation over $5\times5$ nested cross-validation. The network is trained on a subset of 47 bpMRI scans from the training dataset and its output zonal segmentations are used to construct and apply the anatomical priors (as detailed in \hyperref[intro]{Section 1}). Special care is taken to ensure mutually exclusive patients between the training, validation and testing datasets.

\paragraph{Experiments} Adjusting the value of $\mu$ can lead to remarkably different priors, as seen in \hyperref[fig1]{Fig. 1(\textit{a-c}}). We test three different priors, switching the value of $\mu$ between 0.00, 0.01 and 0.33, to investigate the range of its impact on csPCa detection. Based on our observations in previous work \cite{ZonalPrior}, we opt for an early fusion of the probabilistic priori, where each variant of $P$ is stacked as an additional channel in the input image volume  (refer to \hyperref[fig1]{Fig. 1(\textit{d}})) via separate turns. Candidate CNN models include 3D adaptations of the stand-alone U-Net \cite{3DUNet}, an equivalent network composed of \textit{Squeeze-and-Excitation} residual blocks \cite{SE} termed U-SEResNet, the UNet++ \cite{UNet++} and the Attention U-Net \cite{Atn-Gates} architectures. All models are trained using intensity-normalized (\textit{mean}=0, \textit{stdev}=1), center-cropped ($144$$\times$$144$$\times$$18$) images with $0.5$$\times$$0.5$$\times$$3.6$ mm$^3$ resolution. Minibatch size of 4 is used with an exponentially decaying cyclic learning rate \citep{CLR} oscillating between $10^{-6}$ and $2.5\times10^{-4}$. Focal loss ($\alpha=0.75, \gamma=2.00$) \cite{FocalLoss} is used to counter the 1:153 voxel-level class imbalance \cite{UCLA} in the training dataset, with Adam optimizer \cite{Adam} in backpropagation. Train-time augmentations include horizontal flip, rotation ($-7.5\degree$ to $7.5\degree$), translation ($0$-$5\%$ horizontal/vertical shifts) and scaling ($0$-$5\%$) centered along the axial plane. During inference, we apply test-time augmentations by averaging predictions over the original and horizontally-flipped images.

\section{Results and Discussion}
\label{res}

\begin{figure}[h!]
\centering
\includegraphics[width=1.0\textwidth]{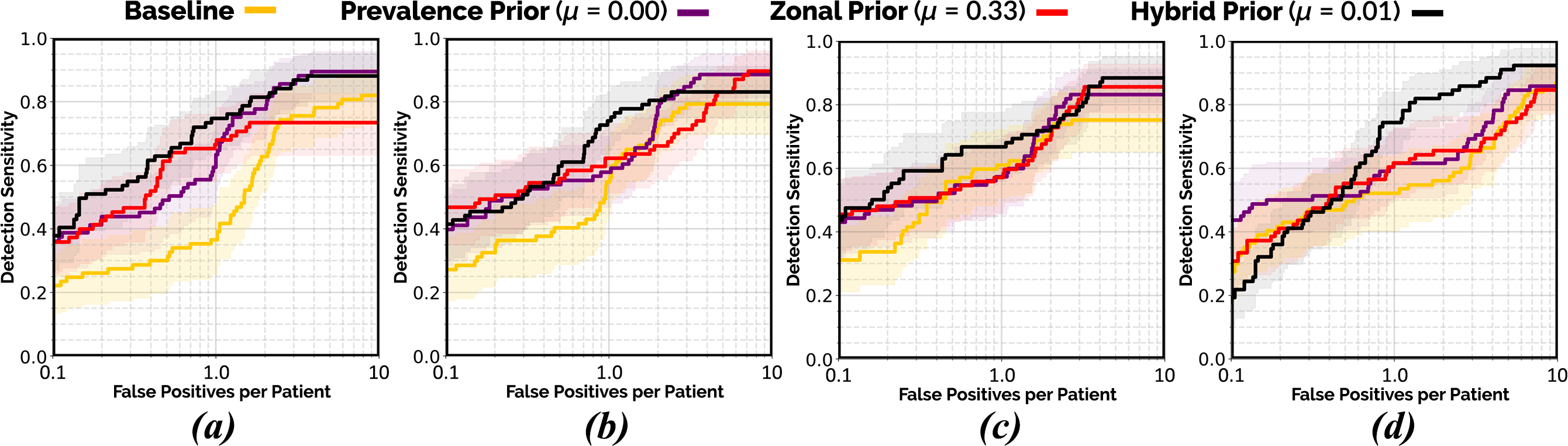}
\caption{Lesion-level \textit{Free-Response Receiver Operating Characteristic} (FROC) analyses of each 3D CNN model paired with different variants of the anatomical prior $P$: \textbf{\textit{(a)}} U-Net \textbf{\textit{(b)}} U-SEResNet \textbf{\textit{(c)}} UNet++ \textbf{\textit{(d)}} Attention U-Net. Transparent areas indicate the 95\% confidence intervals estimated from 1000 replications of bootstrapping.}
\label{fig2}
\end{figure}

Patient-based diagnosis and lesion-level detection performance on the testing set are noted in \hyperref[tab1]{Table 1} and \hyperref[fig2]{Fig 2}, respectively. For every combination of the 3D CNN models and a variant of the prior $P$, we observe improvements in performance over the baseline. Notably, the hybrid prior, which retains a blend of both csPCa prevalence and zonal priori, shares the highest increases of 7.32--8.70\% in patient-based AUROC. $P$ demonstrates a similar ability to enhance csPCa localization, with an average increase of 1.08 in pAUC between 0.1--10 false positives per patient across all FROC setups.

\begin{table}[h!]
  \label{tab1}
  \renewcommand{\arraystretch}{1.00}
  \caption{Patient-based diagnosis performance of each 3D CNN model paired with different variants of the anatomical prior $P$. Performance scores indicate the mean metric followed by the 95\% confidence interval estimated as twice the standard deviation from 1000 replications of bootstrapping.}
  \centering
  \begin{tabular}{
    p{0.2000\textwidth}>
    {\centering}p{0.1625\textwidth}>
    {\centering}p{0.1625\textwidth}>
    {\centering}p{0.1625\textwidth}>
    {\centering\arraybackslash}p{0.1625\textwidth}}
    \toprule
    \multirow{4}{*}{Architecture} & 
    \multicolumn{4}{c}{\textit{Area Under Receiver Operating Characteristic} (AUROC)}       \\
    \cmidrule(r){2-5} 
    & Baseline         & Prevalence Prior & Zonal Prior & Hybrid Prior \\
    & (without prior)  & ($\mu$ = 0.00)   & ($\mu$ = 0.33) & ($\mu$ = 0.01) \\
    \midrule
    U-Net             \cite{3DUNet}     & $0.690{\scriptstyle \pm 0.079}$ & $0.737{\scriptstyle \pm 0.076}$ & $0.740{\scriptstyle \pm 0.073}$ &  $0.763{\scriptstyle \pm 0.071}$\\
    U-SEResNet        \cite{SE}         & $0.694{\scriptstyle \pm 0.077}$ & $0.732{\scriptstyle \pm 0.077}$ & $0.748{\scriptstyle \pm 0.080}$ &  $0.777{\scriptstyle \pm 0.072}$\\
    UNet++            \cite{UNet++}     & $0.694{\scriptstyle \pm 0.078}$ & $0.734{\scriptstyle \pm 0.080}$ & $0.752{\scriptstyle \pm 0.079}$ &  $0.781{\scriptstyle \pm 0.069}$\\
    Attention U-Net   \cite{Atn-Gates}  & $0.711{\scriptstyle \pm 0.078}$ & $0.736{\scriptstyle \pm 0.079}$ & $0.750{\scriptstyle \pm 0.071}$ &  $0.790{\scriptstyle \pm 0.066}$\\
    \bottomrule
  \end{tabular}
\end{table}

\vspace{-3mm}

In this research, we demonstrate how the standardized imaging protocol of prostate bpMRI can be leveraged to construct explicit anatomical priors, which can subsequently be used to encode clinical priori into state-of-the-art U-Net architectures. By doing so, we are able to provide a higher degree of train-time supervision and boost overall model performance in csPCa detection, even in the presence of a limited training dataset with inaccurate annotations. In future study, we aim to investigate the prospects of integrating our proposed prior in the presence of larger training datasets, as well as quantitatively deduce its capacity to guide model generalization to histologically-confirmed testing cases beyond the radiologically-estimated training annotations.


\section*{Broader Impact}

Prostate cancer is one of the most prevalent cancers in men worldwide \cite{PCaStat2019}. In the absence of experienced radiologists, its multifocality, morphological heterogeneity and strong resemblance to numerous non-malignant conditions in MR imaging, can lead to low inter-reader agreement ($<50\%$) and sub-optimal interpretation \cite{mpMRI,LimitedPIRADS2,LimitedPIRADS3}. The development of automated, reliable detection algorithms has therefore become an important research focus in medical image computing, offering the potential to support radiologists with consistent quantitative analysis in order to improve their diagnostic accuracy, and in turn, minimize unnecessary biopsies in patients \cite{UnnBiop,RadPRvsDL}. 

Data scarcity and inaccurate annotations are frequent challenges in the medical domain, where they hinder the ability of CNN models to capture a complete, visual representation of the target class(es). Thus, we look towards leveraging the breadth of clinical knowledge established in the field, well beyond the training dataset, to compensate for these limitations. The promising results of this study verifies and further motivates the ongoing development of state-of-the-art techniques to incorporate clinical priori into CNN architectures, as an effective and practical solution to improve overall performance.

Population priors for prostate cancer can be susceptible to biases that indicate asymmetrical prevalence. For instance, the computed prior may exhibit a relatively higher response on one side (left/right), stemming from an imbalanced spatial distribution of the \textit{malignant} lesions sampled for the training dataset. We strongly recommend adequate train-time augmentations (as detailed in \hyperref[exp]{Section 2}) to mitigate this challenge.

\begin{ack}
The authors would like to acknowledge the contributions of Maarten de Rooij and Ilse Slootweg from Radboud University  Medical Center during the annotation of fully delineated masks of prostate cancer for every bpMRI scan used in this study. This research is supported in part by the European Union H2020: ProCAncer-I project (EU grant 952159). Anindo Saha is supported by the Erasmus+: EMJMD scholarship in Medical Imaging and Applications (MaIA) program.

\end{ack}

\small
\bibliographystyle{unsrtnat}
\bibliography{main.bib}

\newpage
\normalsize
\section*{Appendix: Model Predictions}

\begin{figure}[h!]
\centering

\includegraphics[width=\textwidth]{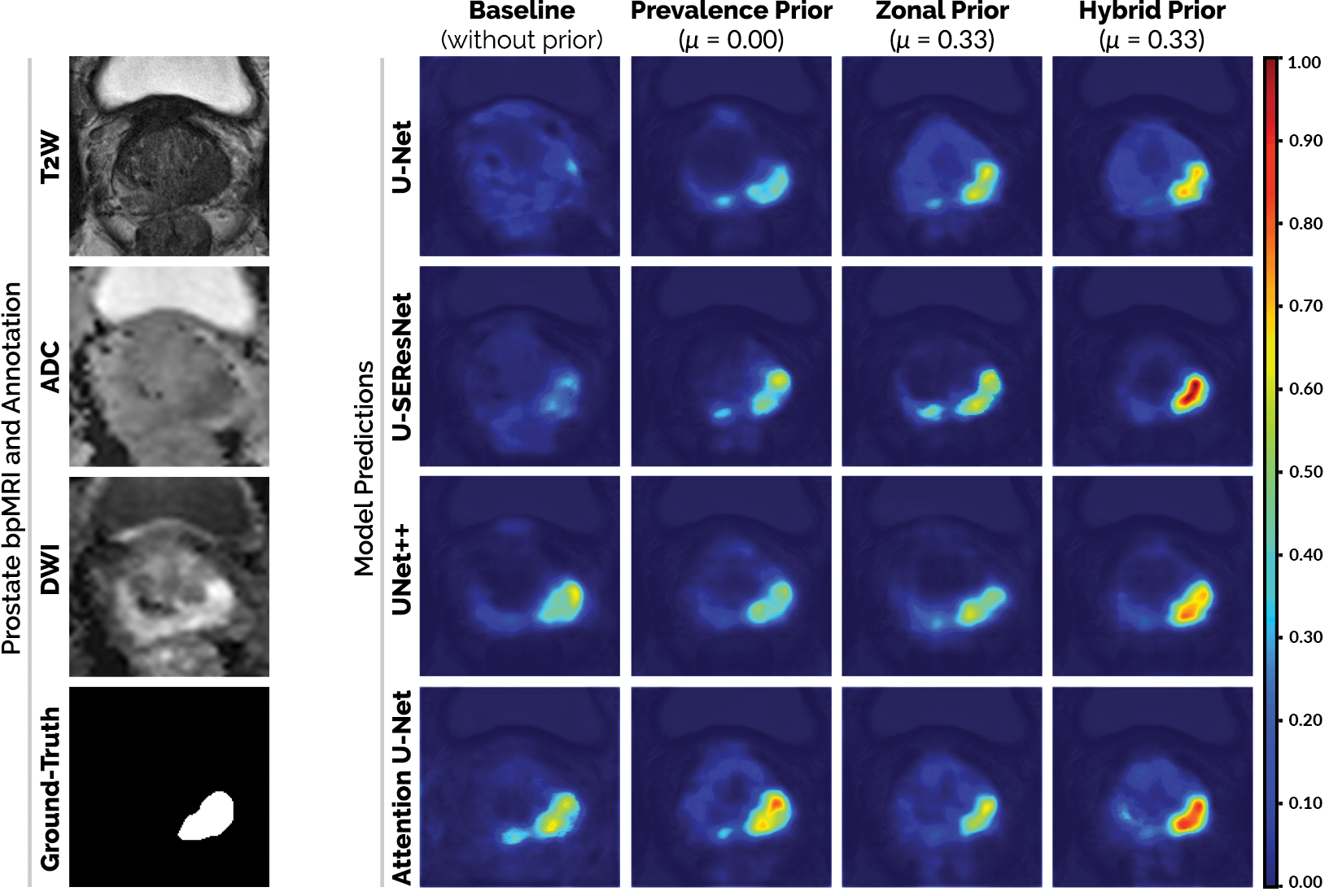}
\textbf{\textit{(a)}}: Histologically-confirmed clinically significant prostate cancer emerging from PZ.

\vspace{4mm}

\includegraphics[width=\textwidth]{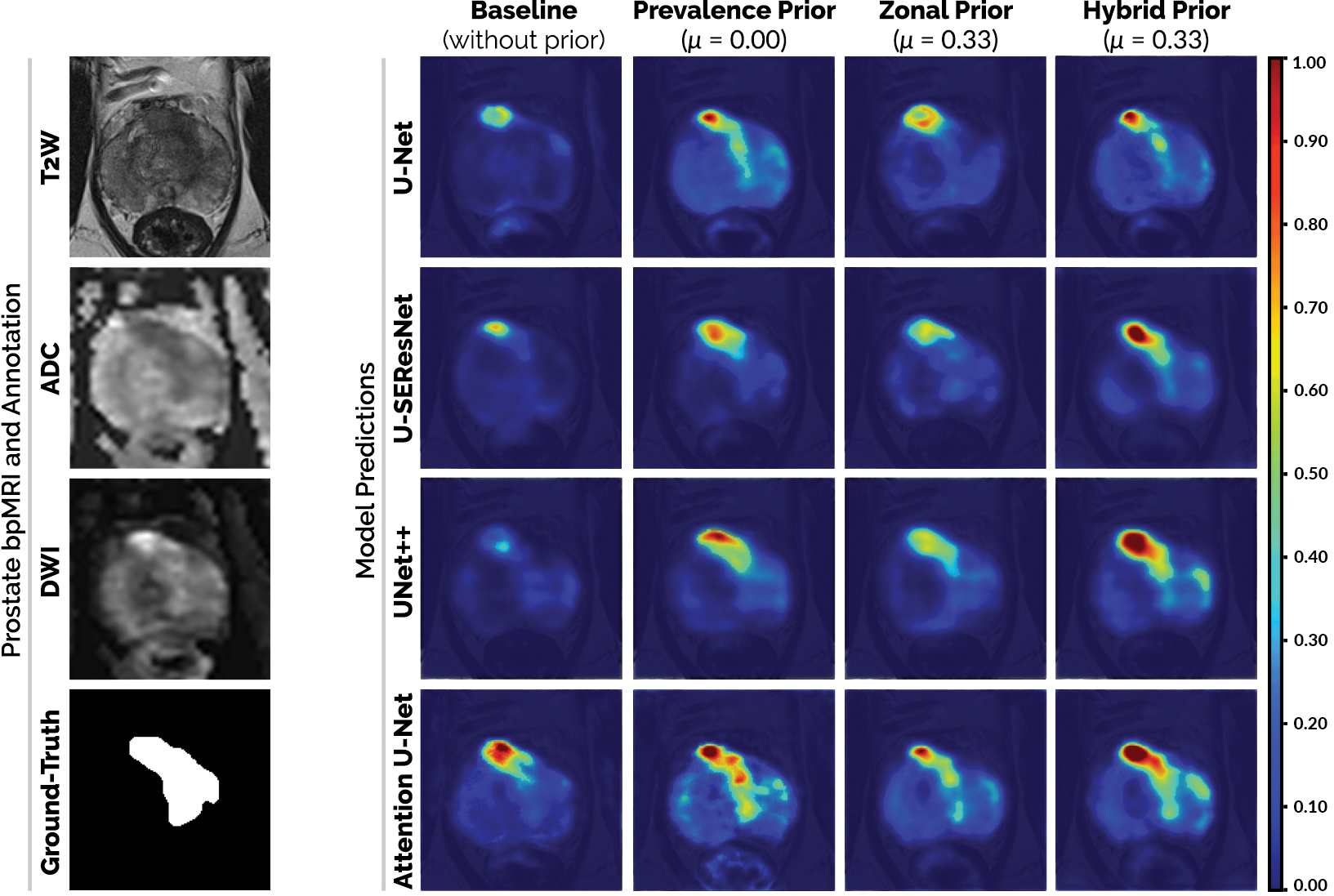}
\textbf{\textit{(b)}}: Histologically-confirmed clinically significant prostate cancer emerging from TZ.

\vspace{2mm}

\caption{Mid-axial bpMRI slice of the prostate gland and its corresponding model predictions (overlaid on T2W images) for two different patient scans in the testing dataset. In each case, the patient is afflicted by a single instance of csPCa localized in a different part of the prostate anatomy.}
\label{fig3}
\end{figure}

\end{document}